\documentclass[12pt]{article}

\begin{document}


\title{ON A RECENTLY PROPOSED RELATION BETWEEN oHS AND ITO SYSTEMS}

\author{ Atalay Karasu\\
{\small Department of Physics , Faculty of Arts and  Sciences}\\
{\small Middle East Technical University , 06531 Ankara-Turkey}}
\date{}
\begin{titlepage}
\maketitle

\begin{abstract}
The bi-Hamiltonian structure of original Hirota-Satsuma system
proposed by Roy based on a modification of the bi-Hamiltonian
structure of Ito system is incorrect.

\end{abstract}
\end{titlepage}

In a recent paper\cite{ROY} a bi-Hamiltonian structure
 for the original Hirota-Satsuma(oHS) system where $a$ is arbitrary
 is proposed and a relation between oHS and Ito systems is introduced
 and a recursion operator is found for the oHS system.
In this note we point out that the oHS system does not in fact admit
a bi-Hamiltonian structure and the relation between the oHS and Ito
systems claimed by the author of \cite{ROY} is actually incorrect.

It is well known that the  original Hirota-Satsuma system (oHS)\cite{HIR}
\begin{eqnarray}
u_{t} & = & a(u_{xxx}+6uu_{x})+2bvv_{x} ,\nonumber\\
v_{t} & = &- v_{xxx}-3uv_{x} \label{a0}
\end{eqnarray}
for all values of $a$ and $b$, possesses three conserved quantities.
\begin{equation}
I_{1}=u,\,\,\,I_{2}=u^{2}+{2 \over 3}bv^{2},\,\,
I_{3}=(1+a)(u^{3}-{1 \over 2}(u_{x})^{2})+b(uv^{2} - (v_{x})^{2}).
\label{a1}
\end{equation}

Later Hirota-Satsuma \cite{HIR1} showed that oHS system has infinitely
many conserved quantities for the choice of $a={1 \over 2}$ and
conjectured that it is completely integrable.
Dodd and Fordy \cite{DOD}
showed that the oHS system admits a  Lax representation only for
this particular value of $a={1 \over 2}$. Also
Aiyer \cite{AIY} proved that the oHS system possesses
a  recursion operator of degree four  only for $a={1 \over 2}$.
The same result has been recently reported
 by the author and G{\" u}rses \cite{GUR}
-\cite{GUR2} in the context of the integrable coupled KdV systems
admitting recursion operators. Wilson \cite{WIL} pointed out that
the oHS system with  $a={1 \over 2}$ belongs to the general
construction of evolutionary equations possessesing Lax-pair
due to Drinfeld and Sokolov \cite{DIR}-\cite{DIR1}.
On the other hand the Ito system \cite{ITO}

\begin{eqnarray}
u_{t} & = & u_{xxx}+6uu_{x}+2vv_{x} ,\nonumber\\
v_{t} & = &2(uv)_{x}. \label{a3}
\end{eqnarray}

admits a bi-Hamiltonian structure
\begin{equation}
B_{II}{\delta {\cal H}_{n}}= B_{I}{\delta {\cal H}_{n+1}} \label{a4}
\end{equation}
where
\begin{eqnarray}
B_{II} \: & = &
\: \left(
\begin{array}{cc}
D^{3}+4uD+2u_{x}
 & 2vD \\
2v_{x}+2vD &
0
\end{array} \; \; \right)\;,
\; B_{I}=
 \left(
\begin{array}{cc}
D
 & 0 \\
0 &
D
\end{array} \; \; \right)\;.
\label{a5}
\end{eqnarray}
\noindent
with the Hamiltonian functionals
\begin{eqnarray}
{\cal H}_{1}[u,v] & = & \int{1 \over 2}(u^{2}+v^{2})dx ,\nonumber\\
{\cal H}_{2}[u,v] & = & \int{1 \over 2}(u^{3}- {1 \over 2}
u_{x}^{2}+uv^{2})dx . \label{a6}
\end{eqnarray}
 The recursion operator arising from a Hamiltonian pair
\begin{equation}
R=B_{II}(B_{I})^{-1}=
 \left(
\begin{array}{cc}
D^{2}+4u+2u_{x}D^{-1}
 & 2v  \\
 2v + 2v_{x}D^{-1}&
  0
\end{array} \; \; \right)\;.
\label{a7}
\end{equation}
is a hereditary operator \cite{GUR1} which gives rise to infinitely many conserved
quantities. The multi-Hamiltonian structure of this system  was studied
by Antonowicz and Fordy \cite{ANT} and by Olver and Rosenau \cite{OLV1}.

At this stage we have the following observations:\\
{\it observation 1}:\\
 The author of \cite{ROY} points out that there is a printing error in the
conserved density $I_{3}$ (instead of $1+a$ there is $a$) obtained
in \cite{HIR}.But this claim is wrong and $I_{3}$ in the ref.\cite{HIR1}
is correct as it is.
 One can check this
either using ${d \over dt}\int{I_{3}}dx=0$
or finding its gradient $\gamma_{3}$
satisfies $ \gamma_{3}^{\prime}[K]+(K^{\prime})^{\dagger}[\gamma_{3}]=0$
and $\gamma_{3}^{\prime} =(\gamma_{3}^{\prime})^{\dagger}$\cite{FOK}.
Here $K$ is the right hand side of the oHS system.\\
{\it observation 2}:\\
 The author of \cite{ROY} claims that the oHS
system can be written as

\begin{equation}
\left(
\begin{array}{c}
u \\ v
\end{array}\right)_{t}
\label{rec}=AB_{1} \left(\begin{array}{c}
{\frac{\delta  {\cal H}_{1}}{\delta u}} \\ {\frac{\delta
{\cal H}_{1}}{\delta v}} \end{array} \right) \;
\end{equation}
with
\begin{equation}
A=
 \left(
\begin{array}{cc}
a(D^{2}+4u+2u_{x}D^{-1})
 & 2v + v_{x}D^{-1} \\
 2v + v_{x}D^{-1}&
-(D^{2}+4u+2u_{x}D^{-1})
\end{array} \; \; \right)\;.
\end{equation}
\noindent
and $B_{I}$ given in (\ref{a5}).
Here ${\cal H}_{1}$ is the Hamiltonian functional of the Ito system.
First we have noticed that ${\cal H}_{1}$ {\it is also a Hamiltonian
functional of the oHS system} ($b=3/2)$. Second, the operator $AB_{1}$
does not satisfy the Jacobi identity \cite{OLV}. Therefore it is not a
Hamiltonian operator for the oHS system
although it is a skew-symmetric operator. Furthermore the author of
\cite{ROY} proposes a bi-Hamiltonian structure for the oHS system
which is based on the bi-Hamiltonian form of Ito system  (\ref{a4}) as

\begin{equation}
AB_{II}{\delta {\cal H}_{n}}= AB_{I}{\delta {\cal H}_{n+1}} \label{a8}
,\,\,\,\,\,n=0,1,2,...
\end{equation}

First of all,{\it the Hamiltonian functional ${\cal H}_{2}$ of the Ito system
 is not a Hamiltonian
functional of the oHS system}. Moreover {\it neither
$AB_{I}$ nor $AB_{II}$ are Hamiltonian operators}. As a result
the expression (\ref{a8})  does not constitute  a bi-Hamiltonian form
 for the oHS system.\\
{\it observation 3}\\
The author of \cite{ROY} claims that an infinite hierarchy of the oHS
system is generated by the recursion operator
\begin{equation}
 {\cal R}=(AB_{II})(AB_{I})^{-1}=ARA^{-1}
\end{equation}
\noindent
where  ${\cal R}$ is the recursion operator for Ito system.
It is easy to see that the given recursion operator ${\cal R}$
 has at most degree two. Therefore it
does not generate the hierarchy of the oHS system. It has been recently
reported that neither oHS nor HS with
 $a={1 \over 2}$ possess a recursion operator of degree two
\cite{GUR}-\cite{GUR2}.

As a conclusion, contrary to the claims made in ref.\cite{ROY},
the oHS system is not a bi-Hamiltonian system .This system is integrable
and admits a bi-Hamiltonian structure only when $a={1 \over 2}$
\cite{FUC}-\cite{OEV}.
 The hierarchy of this system was studied by Levi \cite{LEV}.

 The author would like to thank M. G{\" u}rses for bringing
this recent work on Hirota-Satsuma system  to my attention
and for many valuable discussions.
This work is partially supported by the Scientific and Technical
Research Council of Turkey (TUBITAK).

\end{document}